\documentclass[aps,twocolumn,prl]{revtex4}
\usepackage{amsmath}
\usepackage{graphicx}
\begin{document}

\noindent {\bf Reply to "Comment on 'Fano resonance for Anderson
Impurity Systems' "}

In their Comment, Kolf et al. \cite{kolf05} criticizing  our work
on the Fano resonance for Anderson impurity systems \cite{luo04},
based their argument on the assumption that the Green's function
of d-electron has approximately a Lorentzian form around the Kondo
energy (Eq. (1) in \cite{kolf05}). However, that assumption is
inconsistent with the numerical renormalization group (NRG)
results\cite{costi94}, revealing an asymmetric lineshape of the
impurity quasiparticle peak for systems without particle-hole
symmetry, especially in the mixed valence regime. The asymmetric
lineshape, resulting mainly from the interference between the
Kondo resonance and the broadening impurity level\cite{luo04}, can
strongly affect the low energy behavior of conduction electrons,
in particular the differential conductance measured in STM
experiments, and should not be ignored.

In their Comment, Kolf et al. correctly pointed out that Eq. (8)
in \cite{luo04} overestimates the asymmetry of the impurity
lineshape in the mixed valence regime. However, the error in Eq.
(8) was not caused by Eq. (4) in \cite{luo04} which is rigorous.
It can be derived using the equation of motion method without
invoking Wick's theorem. The error is instead due to an
oversimplification in our approximate expression for
$T_d(\omega)$, Eq. (7) in \cite{luo04}, containing a Kondo
resonance pole and a slowly varying background. The correct low
energy form of $T_d(\omega)$ should be \cite{taylor}:
\begin{equation}
T_d(\omega) \approx \frac{a\,\mbox{e}^{i\delta}}{\omega -
\varepsilon_K + i\Gamma_K} + t_{incoh}, \label{td}
\end{equation}
where $e^{i\delta}$ is the phase factor that was missed in
\cite{luo04}.  In the Kondo limit, $\delta$ $\sim$ $0$ and $a$
$\sim$ $\Gamma_K / \pi \rho_{d,0}$, the above equation reduces to
Eq. (7), while in the mixed valence regime missing of the phase
factor leads to an overestimation of the lineshape asymmetry.
Replacing Eq. (7) in \cite{luo04} with the above equation, the
rest of derivations in \cite{luo04} are still valid. Therefore our
main physical picture and conclusions made in \cite{luo04} remain
unchanged.

Using Eqs. (4-6) in \cite{luo04} and Eq. (\ref{td}) here, we have
reanalyzed the experimental data of Ti/Au and Ti/Ag systems,
assuming $U \rightarrow \infty$ for simplicity. The fitting
parameters are ($n$, $\varepsilon_d$, $\Delta$, $\varepsilon_K$,
$\Gamma_K$, $a$, $\delta$, $q_c$) = ($0.38$, $2.3$, $65.0$,
$-1.9$, $4.0$, $28.2$, $2.7$, $2.0$) for Ti/Au and ($0.53$,
$13.4$, $38.8$, $-1.4$, $5.2$, $144.9$, $3.0$, $1.8$) for Ti/Ag
($\varepsilon_F = 0$ and the unit of energy is meV).  Figure 1
shows that the experimental data can be well described by these
equations. However, after the inclusion of the phase factor, $
\rho_d(\omega)$ cannot be any more expressed in the simplified
form of a Fano resonance as given by Eq. (8) in \cite{luo04}. The
insets show $\rho_d(\omega)$  are  asymmetric, but now without
unphysical dip structure, in qualitative agreement with the NRG
results\cite{costi94}.  The values of the fitting parameters
indicate that both Ti/Au and Ti/Ag systems are in the mixed
valence regime, being consistent with the experimental analysis
and our previous conclusion. Thus their criticism that our
analysis "is conceptually incorrect and the quantitative agreement
of ... is meaningless" is unjustified.
\begin{figure}[h]
\includegraphics[width=8cm]{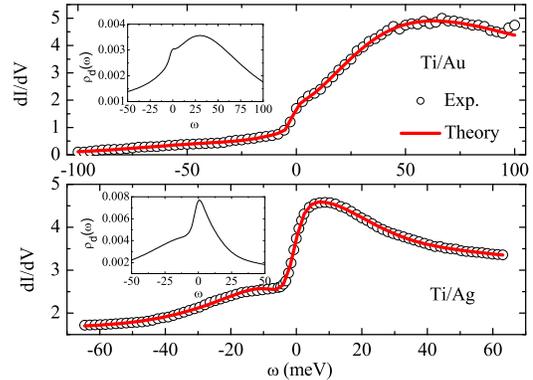}
\vskip-0.5cm \caption{Comparison between theoretical fitting
curves and the STM experimental data for Ti/Au and Ti/Ag. The
inset shows the corresponding impurity density of states.}
\end{figure}

The second comment of \cite{kolf05} is conceptually incorrect.
$\Delta$ and $\Gamma_K$ result from two different physical effects
and represent two different energy scales. They can certainly be
distinguished, at least in the limit $\Gamma_K \ll \Delta$ when
the broadened impurity level can be taken effectively as a
continuum channel and our theory can be applied. In the mixed
valence regime, the fact that one cannot see a sharp peak with
width $\Gamma_K$ does not mean at all the absence of that energy
scale. In the third comment, the authors of Ref. \cite{kolf05}
claimed that the values of $\Delta$ we obtained for Ti/Au and
Ti/Ag are too small. However, they did not give any firm evidence
to support that claim. In fact, as revealed by experiments, the
spectra for different transition metal atoms on Au surface
 behave very differently\cite{jamneala00}. Thus, there is no
reason to expect that the hybridization between a transition metal
atom and conduction electrons should  have the same order of
magnitude.

\vskip 0.1cm
\noindent H. G. Luo, T. Xiang, X. Q. Wang, Z. B. Su, and L. Yu\\
\noindent Institute of Theoretical Physics and Interdisciplinary
Center of Theoretical Studies, Chinese Academy of Sciences, P. O.
Box 2735, Beijing 100080, China

\end{document}